\documentclass[twocolumn,aps,prl,showpacs,floatfix,,superscriptaddress]{revtex4}
\usepackage[latin9]{inputenc}
\usepackage{graphicx}
\usepackage{esint}

\makeatletter
\@ifundefined{textcolor}{}
{%
 \definecolor{BLACK}{gray}{0}
 \definecolor{WHITE}{gray}{1}
 \definecolor{RED}{rgb}{1,0,0}
 \definecolor{GREEN}{rgb}{0,1,0}
 \definecolor{BLUE}{rgb}{0,0,1}
 \definecolor{CYAN}{cmyk}{1,0,0,0}
 \definecolor{MAGENTA}{cmyk}{0,1,0,0}
 \definecolor{YELLOW}{cmyk}{0,0,1,0}
}

\usepackage{bm}\usepackage{float}\usepackage{afterpage}

\makeatother

\begin{document}

\title{The $\pi$ Berry Phase and Topological Excitation in Interacting Boson-Fermion Mixtures}

\author{Huaiming Guo}
\affiliation{Department of Physics, Beihang University, Beijing, 100191, China}

\begin{abstract}
The topological property of boson-fermion mixture in a one-dimensional optical superlattice is studied and the topological insulating phase of interacting boson-fermion mixture characterized by a nontrivial Berry phase is identified. The single-particle and boson-fermion exchanging excitation spectrums are calculated and we identify the boson-fermion exchanging excitation as the gapless topological excitation in the topological phase of the mixture.
The different kinds of excitations are explained explicitly from the competition among the bulk gap, the on-site boson-boson and boson-fermion interactions. The Hamiltonian studied has been partly realized in the state-of-art cold atom experiments, and the results are very relevant to the experiments.
\end{abstract}


\maketitle
\section{Introduction}

The discovery of topological insulators has stimulated great interests in the studies of topological quantum phases \cite{rev1,rev2,rev3,rev4,rev5}.
Besides two- and three- dimensional ones, one-dimensional (1D) topological phases also attract intense recent studies \cite{oned1,oned2,oned3,oned4,oned5,oned6,oned7,oned8}.
1D systems not only exhibit equally rich topological properties, but also have the advantage of reduced complexity. Moreover 1D topological phases are experimentally relevant to the optical and photonic superlattices, in which many aspects of the topological property have been studied, such as: the localized boundary states, the adiabatic pumping, the measurement of the Berry phase, et.al. \cite{op1,op2,cold1,hap1,hap2,hap3,hap4}.

Besides 1D fermion topological phases, 1D boson topological phases have also  been predicted in various models \cite{bose1,bose2,bose3,bose4,bose5}. The studies extend the understanding of topological property to systems with different quantum statistics. Additionally the underlying models, i.e., Bose-Hubbard models, represent one of the simplest systems realized in ultracold atom experiments \cite{bhm1,bhm2}, based on which studying the topological property is an important topic currently. Cold atomic systems have also allowed the interesting realizations of boson-fermion mixtures, which rarely occur in nature. They have been the subject of considerable experimental and theoretical works \cite{mix1,mix2}. It is shown that they exhibit rich physical properties, including many exotic quantum phases \cite{mix3,mix4,mix5}. Naturally an interesting question rises whether the mixtures can possess nontrivial topological properties.

In the paper, the topological property of boson-fermion mixture in a one-dimensional optical superlattice is studied and the topological insulating phase of boson-fermion mixture characterized by a nontrivial Berry phase is identified. For the hardcore-boson case, the topological mixture insulator isn't affected by the boson-fermion interaction. However for the softcore-boson case, the topological mixture insulator is driven to a trivial one by the on-site boson-fermion interaction. We calculate the single-particle and boson-fermion exchanging excitation spectrums to study the bulk-boundary correspondence. In the presence of the on-site boson-fermion interactions, the single-particle excitation is gapped. Nevertheless for small boson-fermion interactions the single-particle excitation energies are still in the gap and the excited particle mainly distributes near the boundaries of the system, which are the remnants of the topological property. It is found that the excitation of boson-fermion exchanging is gapless under open boundary condition (OBC) in the topological mixture insulator. Moreover the bulk gap in such excitation spectrum closes at the topological phase transition point. Thus we identify the boson-fermion exchanging excitation as the topological excitation in the topological phase of the mixture. The two kinds of excitations are explained explicitly from the competition among the bulk gap, the on-site boson-boson and boson-fermion interactions. Since the Hamiltonian studied has been partly realized in the state-of-art cold atom experiments, these results are very relevant to the experiments.

\begin{figure}[htbp]
\centering \includegraphics[width=7.5cm]{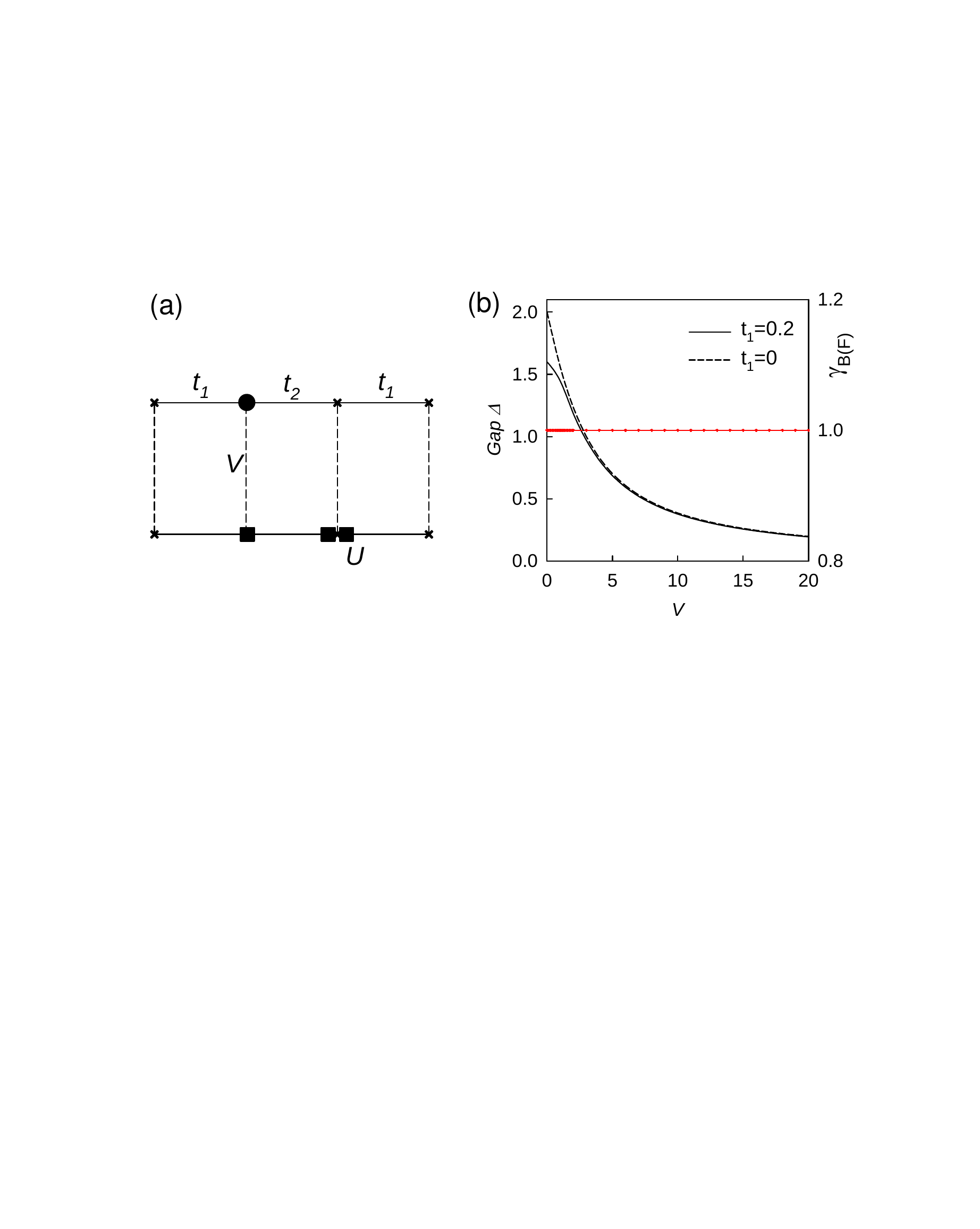} \caption{ (Color online) (a) Schematic illustration of the Hamiltonian Eq.(\ref{eq1}), which consists of one free fermion chain and one interacting bosonic chain coupled by the boson-fermion interaction $V$. $U$ is the strength of the on-site boson-boson interaction. (b) The energy gap $\Delta$ and the Berry phase $\gamma_{B(F)}$ vs $V$.}
\label{fig1}
\end{figure}

\section{Model for Interacting Boson-Fermion Mixture}

We consider the 1D Bose-Fermi Hubbard model described by the Hamiltonian,
\begin{eqnarray}\label{eq1}
 \hat{H}=&-&\sum_i (t_{i,i+1} \hat{f}_i^{\dagger}\hat{f}_{i+1}+H.c.) \\ \nonumber
         &-&\sum_i (t_{i,i+1} \hat{b}_i^{\dagger}\hat{b}_{i+1}+H.c.) \\ \nonumber
 &+&\frac{U}{2}\sum_i \hat{n}_i^{B}(\hat{n}_i^{B}-1)+V\sum_i \hat{n}_i^{B} \hat{n}_i^{F},
\end{eqnarray}
where $\hat{f}_i (\hat{b}_i)$ is the fermionic (bosonic) annihilation and creation operators on the $i$ -th lattice site; the hopping amplitudes $t_{i,i+1}$ are alternating: $t_{i,i+1}=t_1 (t_2)$ for odd (even) $i$ -th site and are assumed identical for bosons and fermions; $\hat{n}_i^{B}=\hat{b}_i^{\dagger}\hat{b}_{i}$ and $\hat{n}_i^{F}=\hat{f}_i^{\dagger}\hat{f}_{i}$ are the number operators for bosons and fermions, respectively; $U, V$ are the strengths of the on-site boson-boson and boson-fermion interactions. Here $t_2=1$ is set as the energy unit.

In the case of noninteracting pure fermions, the Hamiltonian described by Eq.(\ref{eq1}) is Su-Schrieffer-Heeger (SSH) model \cite{ssh}, which writes as,
 \begin{eqnarray}\label{}
 \hat{H}_{SSH}=&-&\sum_i (t_1 \hat{f}_{2i-1}^{\dagger}\hat{f}_{2i}+t_2 \hat{f}_{2i}^{\dagger}\hat{f}_{2i+1}+H.c.). \nonumber
\end{eqnarray}
 In the momentum space, it is $ \hat{{\cal H}}_{SSH}(k)=(t_1+t_2\cos k)\sigma_x+t_2\sin k\sigma_y$. The energy spectrum is $E(k)=\pm\sqrt{(t_2\sin k)^2+(t_1+t_2\cos k)^2}$. The gap of the system is $|t_1-t_2|$. The Hamiltonian has the chiral symmetry $\sigma_z\hat{{\cal H}}_{SSH}(k)\sigma_z=-\hat{{\cal H}}_{SSH}(k)$, so its topological invariant is a winding number $w=-\int_{-\pi}^{\pi}\frac{dk}{2\pi i}\partial_k \textrm{ln} \textrm{Det}(V)$, where $V=t_1+t_2e^{-ik}$ is the upper off-diagonal element of $\hat{{\cal H}}_{SSH}(k)$. Direct calculations give $w=1$ for $t_1<t_2$ while $w=0$ for $t_1>t_2$.
 Alternately since the value of the topological invariant is no more than one, its topological property can also be characterized by the Berry phase, $\gamma=\oint {\cal A}(k) dk$, with the Berry connection ${\cal A}(k)=i\langle u_k|\frac{d}{dk}|u_k\rangle$ and $u_k$ the occupied Bloch state \cite{berry1,berry2}. The Berry phase $\gamma$ mod $2\pi$ takes two values: $\pi$ for a topological insulator and $0$ for a trivial insulator. For the topological phase, there appear a pair of zero modes under OBC, which distribute near the boundaries (also referred as the boundary states). The appearance of the boundary states corresponds to the nontrivial bulk topological invariant, which is known as the bulk-boundary correspondence. The case of $t_1=0$ is special, when the system breaks into independent dimers (two nearest-neighbor sites connected by a bond with the amplitude $t_2$, see Fig.\ref{fig3},\ref{fig6} and \ref{fig7}) and it is in the topological phase with largest gap. Under OBC, the dimer across the boundary is broken into two isolated sites, each of which hosts a zero mode.

 In the case of pure bosons, it has been well studied and the system is in a topological Bose-Mott insulator phase for sufficiently large $U$ at half filling. In the hardcore limit $U=\infty$, the system is exactly solvable by mapping the hardcore-bosons to fermions via the Jordan-Wigner transformation \cite{jordan}. The resulting noninteracting fermionic Hamiltonian is the same as Eq.(\ref{eq1}) of pure fermions, except the term $\hat{b'}_1^{\dagger}\hat{b'}_{L}=-\hat{f'}_1^{\dagger}\hat{f'}_{L} \prod_{n=1}^{L}e^{i\pi \hat{f'}_n^{\dagger}\hat{f'}_n}$, where the hardcore-bosonic operator $\hat{b'}_i^{\dagger}=\hat{f'}_i^{\dagger}\prod_{n=1}^{i-1}e^{i\pi \hat{f'}_i^{\dagger}\hat{f'}_i}$ with $\hat{f'}_i^{\dagger}$ the mapped fermionic operator. Although an additional sign in the hopping across the boundary is induced when the number of hardcore-bosons is even, the topological property is not affected.

In the following, Eq.(\ref{eq1}) is studied using the exact diagonalization (ED) method under a complete basis of the form $|\{n_1^{B},...,n_L^{B}\}_{\alpha}\rangle \otimes |\{n_1^{F},...,n_L^{F}\}_{\beta}\rangle$, where $n_i^B (n_i^{F})$ is the boson (fermion) occupation number of the $i$ -th site and $\alpha (\beta)$ labels all possible compositions with fixed total boson (fermion) occupation number $N_{B(F)}=\sum_i {n}_i^{B(F)}$ \cite{exact}. Here we consider equal filling factors $N_B/L=N_F/L=1/2$ for fermionic and bosonic species.

\section{The hardcore-boson case}

\begin{figure}[htbp]
\centering \includegraphics[width=7.5cm]{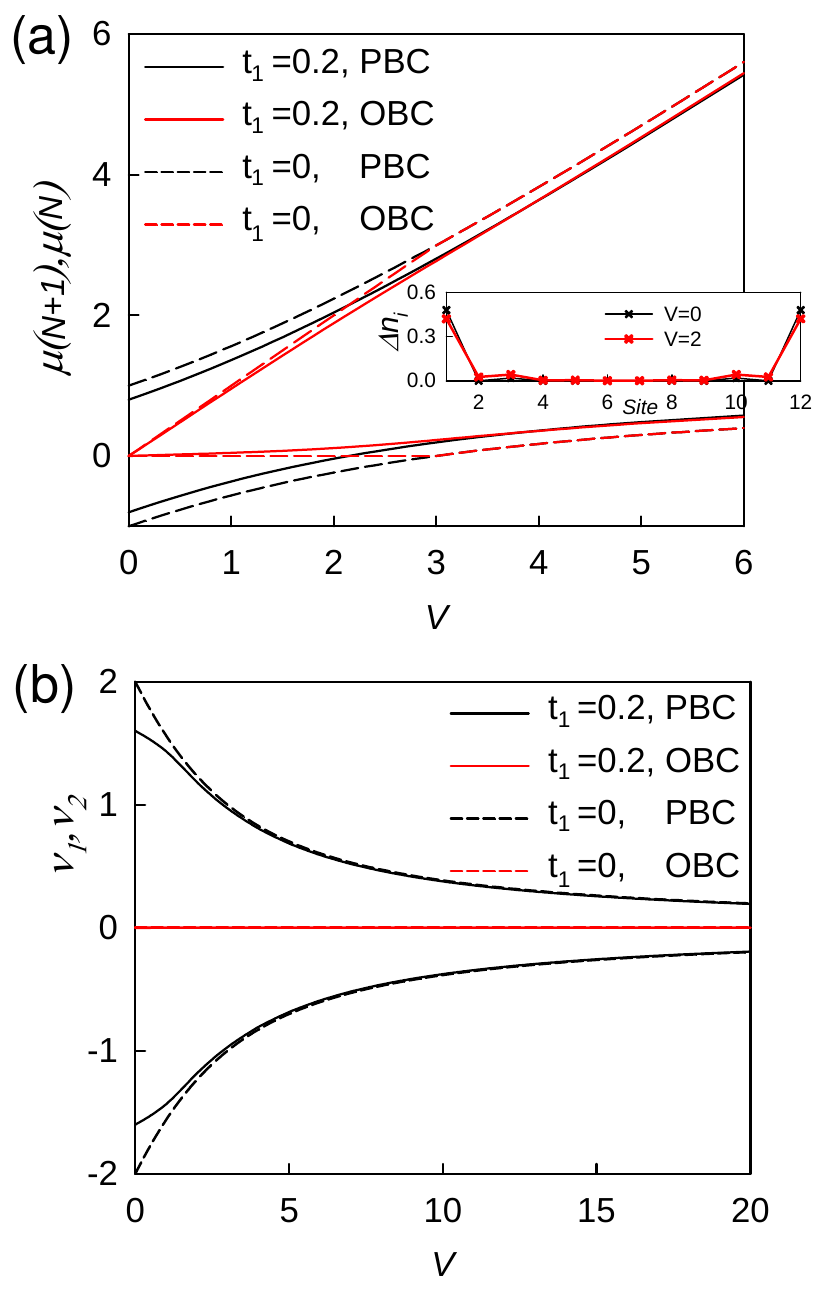} \caption{ (Color online) (a) The chemical potentials of single-particle excitations. (b) The excitation energy of hardcore boson-fermion exchanging. The inset of (a) shows the distribution of the excited boson. The number of lattice sites is $L=8$.}
\label{fig2}
\end{figure}

\begin{figure}[htbp]
\centering \includegraphics[width=7.5cm]{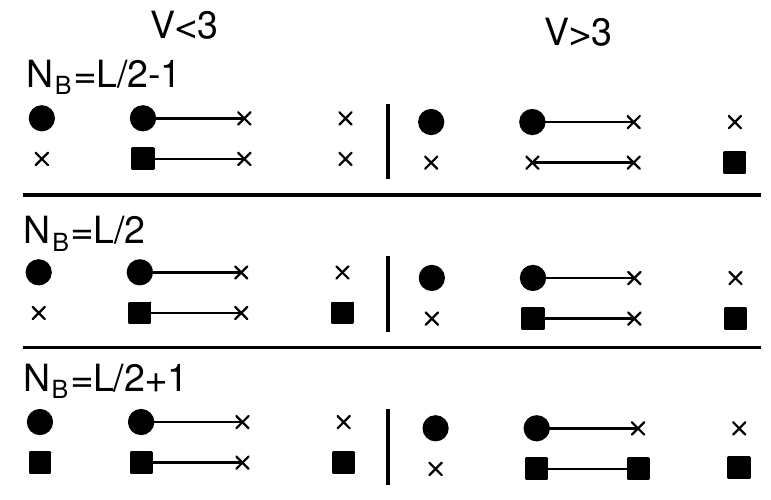} \caption{ Schematic illustration of the configurations of the ground states with different parameters and fillings under OBC in the limit case $t_1=0$, where the solid circles (squares) represent fermions (bosons). In each configuration, the upper (lower) chain is the fermionic (bosonic) part. The crosses represent the sites. Two nearest-neighbor sites connected by a bond form a dimer.}
\label{fig3}
\end{figure}

The hardcore limit is firstly studied, when no multiple occupation of bosons is allowed. We focus on a nontrivial optical superlattice with $t_1<t_2$. At $V=0$, the fermion and boson subsystems are decoupled and the system is a topological mixture insulator. Since the hardcore-bosons can be mapped to fermions, the Hamiltonian Eq.(\ref{eq1}) can be viewed as a fermion model with pseudospin. In the large-$V$ limit, the effective Hamiltonian is the antiferromagnetic Heisenberg model: ${\cal H}_{eff}=\sum_{i}J_{i,i+1} ( {\bf S}_i\cdot {\bf S}_{i+1}-\frac{1}{4}n_i n_{i+1})$,where $n_i=n_{i,1}+n_{i,2}$ is the total particle number on the $i$-th site, $J_{i,i+1}=\frac{4t_{i,i+1}^2}{V^2}$ is the nearest-neighbor coupling and ${\bf S}_{i}=\frac{1}{2}\sum_{\alpha,\beta}f_{i,\alpha}^{\dagger}{\bf \sigma}_{\alpha,\beta} f_{i,\beta}, (\alpha,\beta=1,2)$ is the pseudospin with $f_{i,1}=f_{i}$ the fermion operator, $f_{i,2}=f'_{i}$ the mapped fermion operator, and ${\bf \sigma=(\sigma_x, \sigma_y, \sigma_z})$ the Pauli matrices.

To characterize the topological property of the interacting system ($V\neq 0$), the Berry phase can be defined using the twisted boundary condition, $\gamma=i\oint \langle \psi_{\theta}|\frac{d}{d\theta}|\psi_{\theta}\rangle d\theta$ with $\theta$ the twisted boundary phase, which takes value from $0$ to $2\pi$, and $\psi_{\theta}$ the many-body wave function of $\hat{H}(\theta)$ \cite{berry2,my}. $\hat{H}(\theta)$ is the same as the Hamiltonian described by Eq.(\ref{eq1}) except the hopping terms across the boundary: $\hat{f}^{\dagger}_L\hat{f}_1$ ($\hat{b}^{\dagger}_L\hat{b}_1$), which becomes $\hat{f}^{\dagger}_L\hat{f}_1e^{i\theta}$ ($\hat{b}^{\dagger}_L\hat{b}_1e^{i\theta}$). Since there are two chains in a topological mixture insulator, we use $\gamma_{B(F)}=i\oint \langle \psi_{\theta_{B(F)}}|\frac{d}{d\theta_{B(F)}}|\psi_{\theta_{B(F)}}\rangle d\theta_{B(F)}$ to characterize the topological property of the mixture, where $\theta_{B(F)}$ is the twisted boundary phase acquired only by the bosons (fermions).

The Berry phase $\gamma_{B(F)}$ and the energy gap of the bulk system as a function of the interaction $V$ are calculated. The energy gap is $\Delta=E_{1}-E_{0}$ with $E_{1}, E_{0}$ the eigenenergies of the first-excited state and the ground state of the system under periodic boundary condition (PBC). As shown in Fig.\ref{fig1} (b),  the Berry phase remains $\pi$ all the way, while the energy gap decreases continuously and vanishes at $V=\infty$. The results imply that the topological invariant of the mixture isn't affected by the hardcore boson-fermion interactions.
The nontrivial Berry phase characterizing the topological phase usually corresponds to gapless topological excitations. Next we calculate the single-particle excitation and boson-fermion exchanging excitation to study the bulk-boundary correspondence.
The single-particle excitation gap is defined by $\Delta_c=\mu(N+1)-\mu(N)$, where $\mu(N)$ is the chemical potential computed via $\mu(N)=E_0(N_B,N_F)-E_0(N_B-1,N_F)$ corresponding to single-boson excitation [or $\mu(N)=E_0(N_B,N_F)-E_0(N_B,N_F-1)$ corresponding to single-fermion excitation] with $E_0(N_B,N_F)$ the ground state energy of $N=N_B+N_F$ particles \cite{chemical}. The boson-fermion exchanging excitation gap is defined by $\Delta_s=\nu_2-\nu_1$ with $\nu_2=E_0(L/2-1,L/2+1)-E_0(L/2,L/2)$, $\nu_1=E_0(L/2,L/2)-E_0(L/2+1,L/2-1)$ \cite{oned8}.

We firstly calculate the single-particle  excitation energy. The chemical potentials of adding a particle to half$-1$ and half-filling systems under PBC and OBC are shown in Fig.\ref{fig2} (a) (the excited particle may be hardcore-boson or fermion and the results are the same). As long as $V$ is turned on, the single-particle excitation is gapped. However for small $V$, the chemical potentials are in the bulk gap and the excited particle mainly distributes near the boundaries, implying these excitations are closely related to the boundary states.

The single-particle excitation spectrum in Fig.\ref{fig2}(a) can be understood in the limit case $t_1=0$, when the noninteracting system is deep in the topological phase and the boundary state is totally on the boundary site. We consider the case of hardcore-boson excitations (it is similar for fermion excitations). At half filling and under PBC, each dimer is occupied by one hardcore-boson and one fermion with the energy $E_1=\frac{V}{2}-\frac{\sqrt{V^2+16}}{2}$ (see the Appendix, in which the ground-state energies of different dimers are explicitly calculated). Since the particles behave in a similar way as they do in the SSH model, the nontrivial Berry phase remains. Then we take one hardcore-boson away and the process can happen in any dimer with the excitation energy $\mu(N)=E_1+1$. The process of adding one hardcore-boson to the half-filling system also happens in any dimer and the excitation energy is $\mu(N+1)=V-1-E_1$.

Next we consider the excitation under OBC and start from a system with $L/2-1$ hardcore-bosons. For $V<3$, the ground state under OBC is obtained by cutting off the dimer with only one fermion, generating the configuration shown in Fig.\ref{fig3}. However for $V>3$, it is energetically favorable that the dimer with one hardcore-boson and one fermion is cut off. The energy difference of the above two processes is: $\delta E=-1-E_1$, which is negative for $V>3$.
 Then we add one hardcore-boson to the above system and the number of hardcore-bosons becomes $L/2$. For $V<3$, it just occupies the boundary state and the chemical potential is zero. For $V>3$ it instead occupies the dimer with only one fermion and the chemical potential is the same as the one under PBC. Finally we add one more hardcore-boson. For $V<3$, it occupies the boundary state where one fermion boundary state has already been occupied, thus the chemical potential is $V$. For $V>3$, the added hardcore-boson energetically favors to occupy the dimer with one hardcore-boson and one fermion, and the chemical potentials are the same as those under PBC.  So the results show that the boundary states appear even in the interacting topological phase, but their occupations are determined by a competition between the topological gap and the interaction.

Although the single-particle excitations become gapped in the presence of the interactions, the excitation of hardcore boson-fermion exchanging is gapless, as shown in Fig.\ref{fig2} (b). Though "boson-fermion exchanging" can not be performed directly, but it can be effectly realized by the following way: firstly take away a boson (fermion) from the system, and then add a fermion (boson) into the system. The results can also be understood in the limit case $t_1=0$. For the PBC case, the changing can happen in any dimer and the excitation energies are $\nu_2=-E_1, \nu_1=E_1$. For the OBC case and at half filling, the boundary states are occupied by one hardcore-boson and one fermion, respectively. If the changing happens in the bulk, it is gapped. So it must happen via the boundary states, which is gapless and thus is energetically favored. So the hardcore boson-fermion exchanging excitation represents the gapless topological excitation in the topological phase of the mixture.  In fact it is very similar to the gapless spinor excitation at the boundary of the 1D topological Mott insulator \cite{oned8}.

\section{The softcore-boson case}

\subsection{The ED results}

\begin{figure}[htbp]
\centering \includegraphics[width=7.5cm]{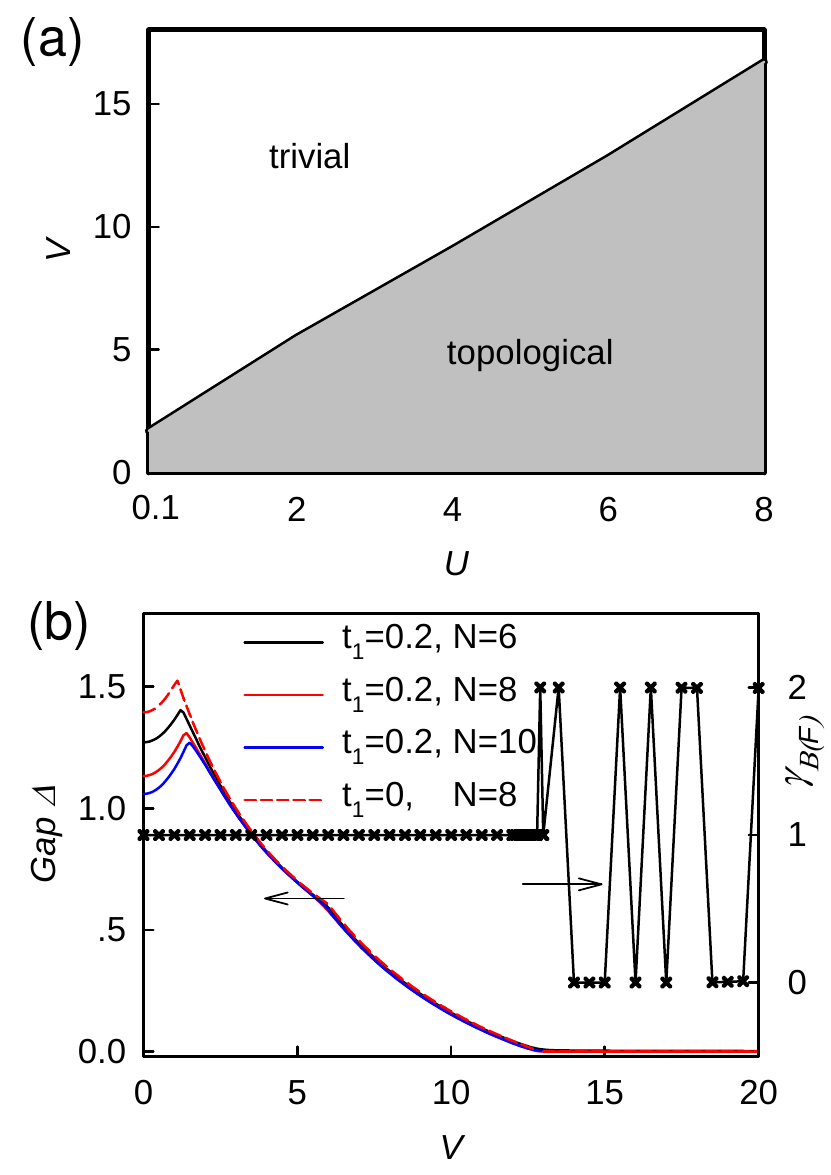} \caption{ (Color online) (a) The phase diagram in the $(U,V)$ plane for the softcore case. (b) The energy gap $\Delta$ and the Berry phase $\gamma_{B(F)}$ as a function of the interaction $V$. In (b), the on-site boson-boson interaction is fixed to $U=6$.}
\label{fig4}
\end{figure}

Next we consider the softcore-boson case and the phase diagram in the $(U,V)$ plane is shown in Fig.\ref{fig4}(a).
The interaction $U$ drives the boson system from a superfluid to a topological Bose-Mott insulator \cite{bose2,jordan,vezzani1,vezzani2}. The phase transition happens at very small strength of the interaction $U$. Then at $V=0$ and sufficiently large $U$, the system consists of a fermionic topological insulator and a bosonic one, which are independent. So the system is in a topological insulating phase of boson-fermion mixture. The further inclusion of the interaction $V$ drives the topological phase into a trivial one.
We perform the calculations at fixed $U$ to show the details.

\begin{figure}[htbp]
\centering \includegraphics[width=7.5cm]{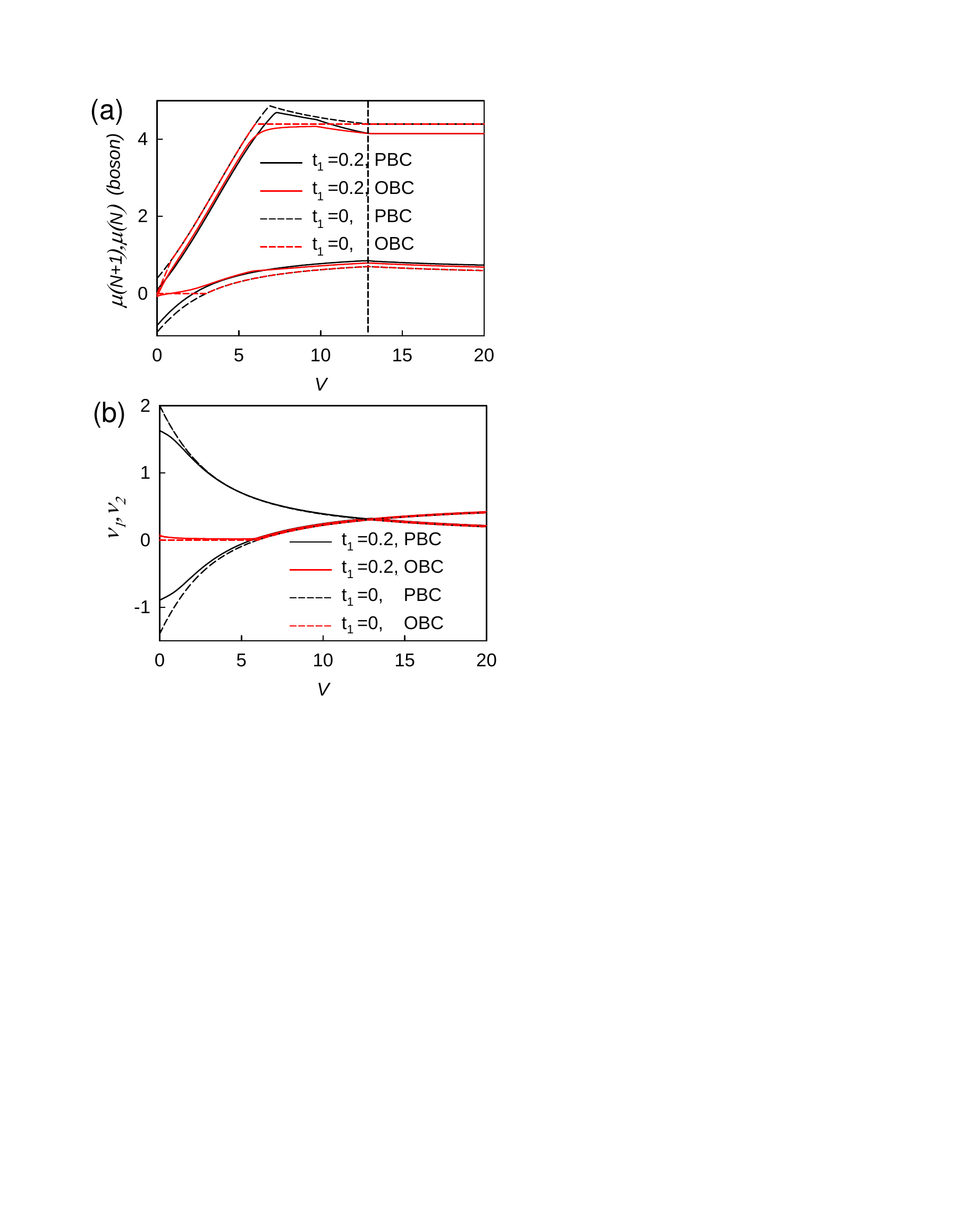} \caption{ (Color online) The different kinds of excitation energies as a function of the interaction $V$: (a) the single-boson excitation; (b) the boson-fermion exchanging excitation. The on-site boson-boson interaction is fixed to $U=6$. The number of lattice sites is $L=8$.}
\label{fig5}
\end{figure}

The energy gap and Berry phase as a function of $V$ are calculated and are shown in Fig.\ref{fig4} (b). A topological phase transition occurs at a critical interaction $V_c$, beyond which the energy gap vanishes and the Berry phase becomes random. To show the topological property, the single-particle excitation energy and the excitation energy of boson-fermion exchanging are calculated.  Similar to the hardcore-boson case, the single-particle excitation is gapped in the presence of $V$, and the chemical potentials are in the bulk gap for small $V$ [see Fig.\ref{fig5} (a)]. However the excitation gaps of boson-fermion exchanging under OBC are gapless for $V<V_c$, and the topological phase transition is accompanied by a closing of bulk gap in the excitation spectrum [see Fig.\ref{fig5} (b)]. So the boson-fermion exchanging excitation represents the topological excitation of the topological phase. The results can be understood from the competition among the bulk gap, the on-site boson-boson and boson-fermion interactions in the limit case $t_1=0$.

\begin{figure}[htbp]
\centering \includegraphics[width=8cm]{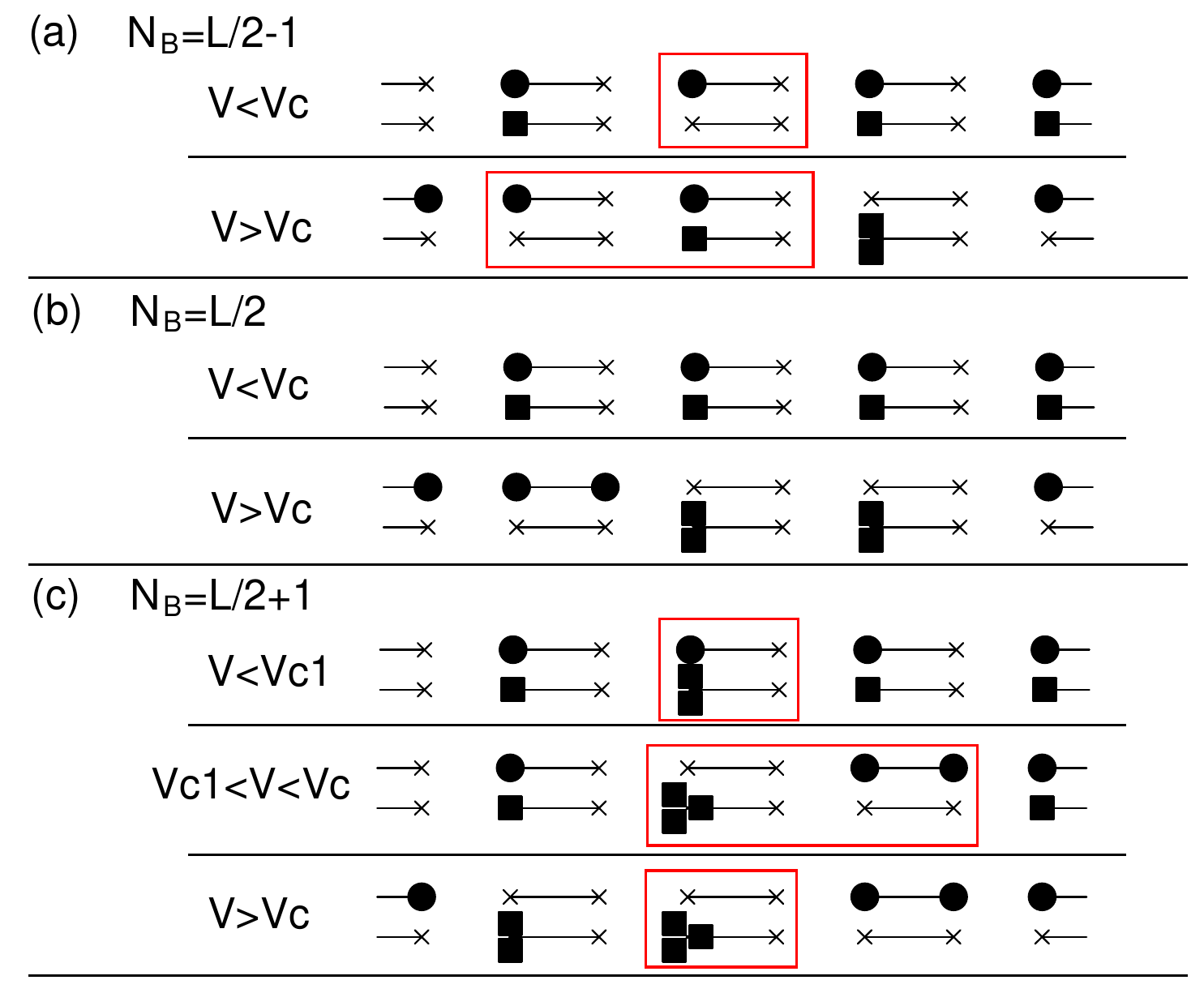} \caption{ (Color online) Schematic illustration of the configurations of the ground states with different parameters and fillings under PBC in the limit case $t_1=0$. Except the first configuration in (b), all other ground state is multi-degenerate and we show only one of them. The number of fermions is $N_F=L/2$. The symbols are the same as those in Fig.\ref{fig3}. The red rectangles mark the dimers which the excitation affects (see the main text).}
\label{fig6}
\end{figure}

\subsection{Analysis from the limit case under PBC}

Firstly a mixture with $N_B=N_F=L/2$ on a periodic chain is considered. For $V<V_c$, each dimer is occupied by one boson and one fermion, and the energy of each dimer is $E_1$ [see Fig.\ref{fig6} (b)]. For $V>V_c$, the phase separation is energetically favored and the ground state is multi-degenerate. The fermions form trivial Mott insulator, while every two bosons occupy a dimer, with the energy $E_2=\frac{U}{2}-\frac{\sqrt{U^2+16}}{2}$. At the critical point, since every two dimers with one fermion and one boson has the same energy as the two dimers with bosons and fermions separated, we have $2E_1=E_2$. The obtained critical interaction $V_c=\frac{(E_2^2-16)}{2E_2}$ ($V_c\simeq 12.9$ for $U=6$) is consistent with the one determined by the Berry phase and the energy gap shown in Fig.\ref{fig4} (b). In the topological phase for $V<V_c$, each dimer is occupied by one boson and one fermion and the particles hop just like the noninteracting case. So its topological invariant is the same as that of the noninteracting case and the system is a topological phase of boson-fermion mixture. Then we take one boson away. For $V<V_c$ the boson can be taken away from any dimer, resulting in a configuration shown in the upper one of Fig.\ref{fig6} (a) and the chemical potential is $\mu(N)=E_1+1$. For $V>V_c$, if the boson is directly taken away from a dimer with two bosons, the resulting dimer with one boson has a energy $-1$. If a fermion in the dimer with two fermions is further removed to the above dimer with one boson, the energy is lowered to $E_1-1$. Thus the resulting configuration [the lower one of Fig.\ref{fig6} (a)] is the ground state. Next we add one boson to the system at half filling. For small $V$ the boson can be directly added to any dimer, generating a dimer with one fermion and two bosons. The energy of such a dimer is $E_3$, so the chemical potential is $\mu(N+1)=E_3-E_1$. At a critical interaction $V_{c1}$, the first two configurations shown in Fig.\ref{fig6} (c) have the same energies, i.e. $E_3+E_1=E_4$ with $E_4$ the energy of the dimer with three bosons, through which $V_{c1}$ is determined ($V_{c1}\simeq 6.87$ for $U=6$).  Then for $V_{c1}<V<V_c$, the second configuration shown in Fig.\ref{fig6} (c) becomes the ground state and the chemical potential is $\mu(N+1)=E_4-2E_1$. For $V>V_c$ the boson can be added to any dimer with two bosons and the excitation energy is $E_4-E_2$.

\subsection{Analysis from the limit case under OBC}

\begin{figure}[htbp]
\centering \includegraphics[width=8cm]{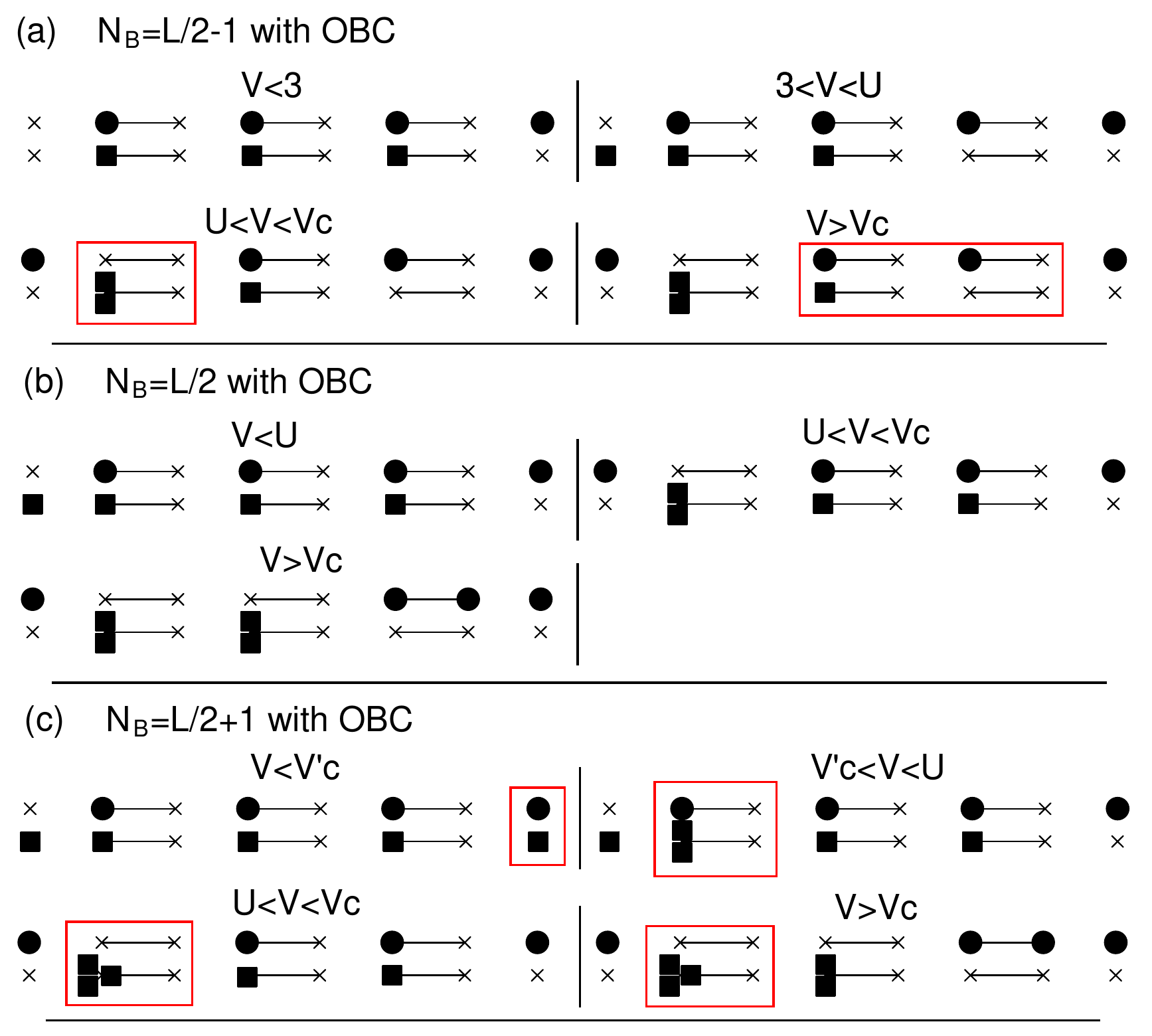} \caption{(Color online) Schematic illustration of the configurations of the ground states with different parameters and fillings under OBC in the limit case $t_1=0$. The number of fermions is $N_F=L/2$. The symbols are the same as those in Fig.\ref{fig6}.}
\label{fig7}
\end{figure}

Next we study the ground state and the excitation energy under OBC. The OBC is obtained by cutting off one dimer, but the redistribution of the particles may happen to let the system in the ground state. We start from the case $N_B=L/2-1$. For $V<U$, the situation is the same as the hardcore case [see Fig.\ref{fig7}(a) and Fig.\ref{fig3}, in which the upper two figures have the same configurations]. For $U<V<V_c$, after cutting off the dimer with one fermion and one boson, the boson on the boundary site moves to one dimer, while the fermion in the dimer moves to the boundary site. The energy cost of such a process is $\delta E =E_2-E_1<0$, which is energetically favored. For $V>V_c$, the system is in phase separation except two dimers in which there are two fermions and one boson. The energetically favored configuration is the one with one fermion in one dimer and the other two particles in the other dimer, as shown in Fig.\ref{fig7} (a). We then add one boson, resulting in a system with $N_B=L/2$. For $V<3$, the added boson just occupies the boundary state, which corresponds to the zero-energy excitation in the single-particle excitation spectrum shown in Fig.\ref{fig7}(b). For $3<V<V_c$, the added boson occupy the dimer with only one fermion and the chemical potential is $\mu(N)=E_1+1$. For $V>V_c$, the boson is added to the dimer with one fermion and one boson, meanwhile the fermion in the dimer moves to the dimer with one fermion. The chemical potential is $\mu(N)=E_2-E_1+1$. The chemical potentials of the above two cases are the same as those under PBC, as shown in Fig.\ref{fig5}(a). Finally we further add one boson and the number of bosons becomes $N_B=L/2+1$. For small $V$, the added boson occupies one of the boundary state with the excitation energy $V$. From a critical interaction $V'_c$ \cite{note}, the boson begins to occupy the dimer with one fermion and one boson. The excitation energy is $\mu(N+1)=E_3-E_1$, which is the same as the bulk one. For $U<V<V_c$, the boson occupies the dimer with two bosons. The excitation energy is $\mu(N+1)=E_4-E_2$, which is lower compared to the bulk one. For $V>V_c$, the added boson enters one dimer with two bosons and the excitation energy is also $\mu(N+1)=E_4-E_2$, which is the bulk one. The single-particle excitation energy of fermion can be analyzed similarly. Compared to the hardcore case, the softcore bosons can condense and the occupation of the boundary is also determined by additional competitions with on-site boson-boson interaction $U$.

 The excitation energy of boson-fermion exchanging at half filling can be analyzed similarly. For the case with PBC, the changing happens in any dimer [see Fig.\ref{fig6}(b)] and the excitation energies (see the definitions in the previous section) are: $\nu_1=E_1-E_2, \nu_2=-E_1$ for $V<V_c$ and $\nu_1=-E_1, \nu_2=E_1-E_2$ for $V>V_c$. Since $2E_1=E_2$ holds at $V_c$, the excitation energy has a closing here. Next we consider the cases under OBC. For $V<U$ the changing happens on the boundary, so the excitation energies are zero. For $V>U$ the changing happens in the bulk. The excitation energy in the region $U<V<V_c$ are $\nu_1=E_1-E_2, \nu_2=E_1-E_2$, which are equal and are the same with one branch under PBC. For $V>V_c$, $\nu_1=-E_1, \nu_2=E_1-E_2$, which are the same with those under PBC. It is notable that for $U<V<V_c$ although the excitation gap of boson-fermion exchanging is gapless, the excitation happens in the bulk.

\section{Conclusions}

The topological property of boson-fermion mixture in a one-dimensional optical superlattice with alternating hopping amplitudes is studied. The topological insulating phase of boson-fermion mixture is identified, which is characterized by a nontrivial Berry phase. The single-particle and boson-fermion exchanging excitation spectrums are calculated and the boson-fermion exchanging excitation is identified as the gapless topological excitation of the topological phase of the mixture.
The two kinds of excitations are explained explicitly from the competition among the bulk gap, the on-site boson-boson and boson-fermion interactions.
The Hamiltonian studied has been partly realized in the state-of-art cold atom experiments and the topological property has been studied. Besides, the recent techniques of controlling optical potential at the single-site level make the studies of the boundary states available \cite{cold1,exp}. So the results presented here are very possibly studied experimentally in the future.

\section*{Acknowledgments}
I thank S. Chen and S. Q. Shen for helpful discussions. This work was supported by NSFC under Grants No.11274032 and No. 11104189.

\appendix
\section{The ground-state energy of the dimers with different numbers of particles}
\begin{figure}[htbp]
\centering \includegraphics[width=7cm]{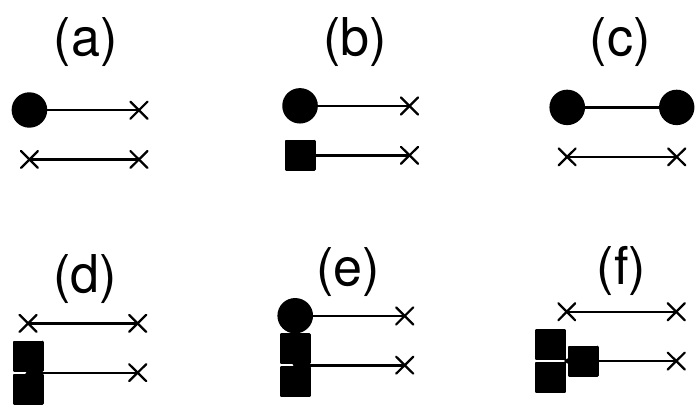} \caption{The dimers with different numbers of particles: (a) $N_{F}=1$; (b) $N_{F}=1, N_{B}=1$; (c) $N_{F}=2$; (d) $N_{B}=2$; (e) $N_{F}=1, N_{B}=2$; (f) $N_{B}=3$. The symbols are the same as those in Fig.\ref{fig6}.}
\label{fig8}
\end{figure}

We explicitly calculate the ground-state energy of the dimers with different numbers of particles. The basis we use has the general form $|i,j;k,l\rangle$ with $i,j(k,l)$ the fermion (boson) numbers on the two sites of the dimer (if the dimer is occupied by only fermions or bosons, the basis becomes $|i,j\rangle$ with $i,j$ the fermion or boson numbers).

For the dimer shown in Fig.\ref{fig8} (a), the basis is: $(|1,0\rangle,|0,1\rangle)$, under which the Hamiltonian matrix is,
 \begin{eqnarray}\label{}
 H_a=\left(
       \begin{array}{cc}
         0 & -t_2 \\
         -t_2 & 0 \\
       \end{array}
     \right). \nonumber
\end{eqnarray}
Its ground-state energy is $-t_2$. The case with one boson in the dimer is the same.

For the dimer shown in Fig.\ref{fig8} (b), the basis is: $(|1,0;0,1\rangle,|0,1;1,0\rangle, |1,0;1,0\rangle, |0,1;0,1\rangle)$, under which the Hamiltonian matrix is,
 \begin{eqnarray}\label{}
 H_b=\left(
       \begin{array}{cccc}
         0 & 0 & -t_2 & -t_2 \\
         0 & 0 & -t_2 & -t_2 \\
         -t_2 & -t_2 & V & 0 \\
         -t_2 & -t_2 & 0 & V \\
       \end{array}
     \right)
 . \nonumber
\end{eqnarray}
Its ground-state energy is $E_1=\frac{V}{2}-\frac{\sqrt{V^2+16t_2^2}}{2}$.

For the dimer shown in Fig.\ref{fig8} (c), the basis is: $(|1,1\rangle)$. The Hamiltonian matrix is $0$, so the energy of the dimer is $0$. If there is one more boson in the dimer, we use the basis: $(|1,1;1,0\rangle, |1,1;0,1\rangle)$ and the Hamiltonian matrix writes as,
\begin{eqnarray}\label{}
 H_c=\left(
       \begin{array}{cc}
         V & -t_2 \\
         -t_2 & V \\
       \end{array}
     \right). \nonumber
\end{eqnarray}
Its ground-state energy is $V-t_2$.

For the dimer shown in Fig.\ref{fig8} (d), the basis is: $(|2,0\rangle, |1,1\rangle, |0,2\rangle)$, under which the Hamiltonian matrix is,
 \begin{eqnarray}\label{}
 H_d=\left(
       \begin{array}{ccc}
         U & -\sqrt{2}t_2 & 0 \\
         -\sqrt{2}t_2 & 0 & -\sqrt{2}t_2 \\
         0 & -\sqrt{2}t_2 & U \\
       \end{array}
     \right)
 . \nonumber
\end{eqnarray}
Its ground-state energy is $E_1=\frac{V}{2}-\frac{\sqrt{U^2+16t_2^2}}{2}$.

For the dimer shown in Fig.\ref{fig8} (d), the basis is: $(|1,0;2,0\rangle, |1,0;1,1\rangle, |1,0;0,2\rangle, |0,1;2,0\rangle, |0,1;1,1\rangle, |0,1;0,2\rangle)$, under which the Hamiltonian matrix is,
 \begin{eqnarray}\label{}
    H_e=\left(
      \begin{array}{cccccc}
        2V+U & -\sqrt{2}t_2 & 0 & -t_2 & 0 & 0 \\
        -\sqrt{2}t_2 & V & -\sqrt{2}t_2 & 0 & -t_2 & 0 \\
        0 & -\sqrt{2}t_2 & U & 0 & 0 & -t_2 \\
        -t_2 & 0 & 0 & U & -\sqrt{2}t_2 & 0 \\
        0 & -t_2 & 0 & -\sqrt{2}t_2 & V & -\sqrt{2}t_2 \\
        0 & 0 & -t_2 & 0 & -\sqrt{2}t_2 & 2V+U \\
      \end{array}
    \right). \nonumber
\end{eqnarray}
Its eigenvalues generally can not be expressed in simple analytical forms, but can be obtained numerically. Its ground-state energy is denoted as $E_3$.

For the dimer shown in Fig.\ref{fig8} (f), the basis is: $(|3,0\rangle, |2,1\rangle, |1,2\rangle, |0,3\rangle)$, under which the Hamiltonian matrix is,
 \begin{eqnarray}\label{}
H_f=\left(
   \begin{array}{cccc}
     3U & -\sqrt{3}t_2 & 0 & 0 \\
     -\sqrt{3}t_2 & U &-2t_2 & 0 \\
     0 & -2t_2 & U & -\sqrt{3}t_2 \\
     0 & 0 & -\sqrt{3}t_2 & 3U \\
   \end{array}
 \right). \nonumber
 \end{eqnarray}
 Its eigenvalues are $2U\pm \sqrt{U^2+ 2Ut_1+4t_1^2}-t_1, 2U\pm \sqrt{U^2-2Ut_1+4t_1^2}+t_1$ and the lowest one corresponding to the ground state is $E_4=2U- \sqrt{U^2+ 2Ut_1+4t_1^2}-t_1$.


\begin{thebibliography}{10}

\bibitem{rev1} J. E. Moore, Nature \textbf{464}, 194 (2010).

\bibitem{rev2} M. Z. Hasan and C. L. Kane, \rmp \textbf{82}, 3045
(2010).

\bibitem{rev3} X. L. Qi and S. C. Zhang, \rmp \textbf{83}, 1057
(2011).

\bibitem{rev4} S. Q. Shen, Topological Insulators (Springer, Berlin,
2012).

\bibitem{rev5} M. Hohenadler and F. F. Assaad, J. Phys.: Condens. Matter {\bf 25}, 143201 (2013).

\bibitem{oned1} A. P. Schnyder, S. Ryu, A. Furusaki and A. W. W.
Ludwig, \prb 78, 195125 (2008).

\bibitem{oned2} X.Chen, Z.-C. Gu, Z.-X. Liu and X.-G. Wen, Science {\bf 338}, 1604 (2012).

\bibitem{oned3} L. Fidkowski and A. Kitaev, \prb {\bf 81}, 134509 (2010).

\bibitem{oned4} A. M. Turner, F. Pollmann and E. Berg, \prb {\bf 83} 075102 (2011).

\bibitem{oned5} E. Tang and X.-G. Wen, \prl {\bf 109}, 096403 (2012).

\bibitem{oned6} J. T. Song and E Prodan, \prb {\bf 89}, 224203 (2014).

\bibitem{oned7} D. Sticlet, L. Seabra, F. Pollmann and J. Cayssol, \prb {\bf 89} 115430 (2014).

\bibitem{oned8} T. Yoshida, R. Peters, S. Fujimoto and N. Kawakami, \prl {\bf  112}, 196404 (2014).

\bibitem{op1}  Y.E. Kraus, Y. Lahini, Z. Ringel, M. Verbin, and O. Zilberberg, \prl {\bf 109}, 106402 (2012).

\bibitem{op2} M. Verbin, O. Zilberberg, Y. Kraus, Y. Lahini, and Y. Silberberg, \prl {\bf 110} 076403 (2013).

\bibitem{cold1} M. Atala, M. Aidelsburger, J. T. Barreiro, D. Abanin, T. Kitagawa, E. Demler and I. Bloch, Nature Phys. {\bf 9}, 795 (2013).

\bibitem{hap1} L. J. Lang, X. Cai and S. Chen, \prl {\bf 108}, 220401 (2012).

\bibitem{hap2} S. Ganeshan, K. Sun and S.D. Sarma, \prl {\bf 110} 180403 (2013).

\bibitem{hap3} H.-M. Guo and S. Chen, \prb {\bf 91} 041402 (2015).

\bibitem{hap4} Z. Xu, S. Chen, \prb {\bf 88}, 045110 (2013).

\bibitem{bose1} S.-L. Zhu, Z.-D. Wang, Y.-H. Chan and L.-M. Duan, \prl {\bf 110}, 075303 (2013).

\bibitem{bose2} F. Grusdt, M. Honing and M. Fleischhauer, \prl {110}, 260405 (2013).

\bibitem{bose3} X. Deng, L. Santos, \pra {\bf 89}, 033632 (2014).

\bibitem{bose4}F. Matsuda, M. Tezuka, N. Kawakami, Journal of the Physical Society of Japan, {\bf 83}, 083707 (2014).

\bibitem{bose5} T. Li, H. Guo, S. Chen, and S.-Q Shen, \prb {\bf 91}, 134101 (2015).

\bibitem{bhm1} M. Greiner, O. Mandel, T. Esslinger, T. W. Hansch and I. Bloch, Nature {\bf 415}, 39 (2002).
\bibitem{bhm2} I. Bloch, J. Dalibard, W. Zwerger W, \rmp {\bf 80}, 885, 2008.

\bibitem{mix1} S. Giorgini, L.P. Pitaevskii, S. Stringari, \rmp {\bf 80}, 1215 (2008).

\bibitem{mix2} M.A. Cazalilla, R. Citro, T. Giamarchi T, et al, \rmp, {\bf 83}, 1405 (2011).


\bibitem{mix3} R. Roth, K. Burnett, \pra {\bf 69}, 021601 (2004).

\bibitem{mix4} L. Pollet, M. Troyer, K. Van Houcke, et al, \prl, {\bf 96}, 190402 (2006).

\bibitem{mix5} F. Hebert, G.G.Batrouni, X. Roy, and V.G. Rousseau, \prb {\bf 78}, 184505 (2008).

\bibitem{ssh}
W. P. Su, J. R. Schrieffer, and A. J. Heeger, \prb {\bf 22}, 2099 (1980).

\bibitem{berry1} R. Resta, Rev. Mod. Phys. \textbf{66}, 899 (1994).

\bibitem{berry2} D. Xiao, M.C. Chang, and Q. Niu, \rmp \textbf{82},
1959 (2010).

\bibitem{jordan} V.G. Rousseau, D.P. Arovas, M. Rigol, F. Hebert, G.G. Batrouni, and R.T. Scalettar, \prb {\bf 73}, 174516 (2006).

\bibitem{exact} M. Zhang and R.X. Dong, Eur. J. Phys. {\bf 31}, 591 (2010).

\bibitem{my} H. Guo and S.Q. Shen, \prb {\bf 84}, 195017 (2011); H. M. Guo, Y. Lin, and S. Q. Shen, \prb {\bf 90}, 085413 (2014); H. M. Guo, \pra {\bf 86}, 055604 (2012).

\bibitem{vezzani1} P. Buonsante, V. Penna, and A. Vezzani, \pra {\bf 70}, 061603(R) (2004).

\bibitem{vezzani2} P. Buonsante, V. Penna, and A. Vezzani, \pra {\bf 72}, 031602(R) (2005).

\bibitem{chemical} G. Roux, T. Barthel, I. P. McCulloch, C. Kollath, U. Schollwock, and T. Giamarchi, \pra {\bf 78}, 023628 (2008).



\bibitem{note} $V'_c$ is determined by the relation $V=E_3-E_1$ and $V'_c\simeq 0.86$ for $t_2=0$.

\bibitem{exp} C. Weitenberg, M. Endres, J. F. Sherson, M. Cheneau, P. Schaub, T. Fukuhara, I. Bloch, and S. Kuhr, Nature {\bf 471}, 319 (2011).	

\end{thebibliography}
\end{document}